\documentclass[aps,pre,reprint,onecolumn,12pt,eprint,raggedbottom,citeautoscript]{revtex4-2} %

\usepackage[T1]{fontenc}
\usepackage[english]{babel}
\usepackage{microtype, setspace}
\usepackage{amsmath, amssymb, amsthm, amsfonts, mathtools, mathrsfs}
\usepackage{csquotes, enumitem, moreenum}
\usepackage[table, dvipsnames]{xcolor}
\usepackage{tikz}
\usepackage{soul}
\usepackage{hyperref}
\usepackage{xurl}
\usepackage[capitalize]{cleveref}
\hypersetup{colorlinks=true, allcolors=blue!50!black, breaklinks=true}

\pdfoutput=1
\allowdisplaybreaks
\crefname{section}{Sec.}{Sects.}

\newcommand{\ce}{\coloneqq}
\newcommand{\ec}{\eqqcolon}
\newcommand{\wt}[1]{\widetilde{#1}}

\newcommand{\N}{\mathbb{N}}

\newcommand{\R}{\mathbb{R}}
\newcommand{\C}{\mathbb{C}}
\newcommand{\SFR}{B}
\newcommand{\SFS}{S}

\newcommand{\MD}{\mathcal{D}}
\newcommand{\MF}{\mathcal{F}}
\newcommand{\MH}{\mathcal{H}}
\newcommand{\MO}{\mathcal{O}}
\newcommand{\MSH}{\mathscr{H}}
\newcommand{\MSN}{\mathscr{N}}
\newcommand{\BO}{\mathscr{B}}
\newcommand{\ii}{\mathrm{i}}
\newcommand{\ee}{\mathrm{e}}
\newcommand{\id}{\mathop{}\!\mathrm{Id}}
\newcommand{\diff}{\mathop{}\!\mathrm{d}}
\newcommand{\set}[2][]{#1\{{#2}#1\}}

\newcommand{\norm}[2][]{#1\Vert{#2}#1\Vert}
\newcommand{\ket}[1]{\mathinner{|{#1}\rangle}}
\newcommand{\bra}[1]{\mathinner{\langle{#1}|}}
\newcommand{\braket}[2][]{#1\langle{#2}#1\rangle}
\newcommand{\ketbra}[2]{\ket{#1} \! \bra{#2}}
\DeclareMathOperator{\tr}{tr}

\theoremstyle{remark}
\newtheorem{remark}{Remark}

\newcommand{\FUBaffiliation}{\affiliation{Freie Universität Berlin, Institute of Mathematics, Arnimallee 6, 14195 Berlin, Germany}}

\begin{document}

\title{Open quantum systems and the grand canonical ensemble}

\author{Benedikt M. Reible}
\email{benedikt.reible@fu-berlin.de}

\author{Luigi Delle Site}
\email{luigi.dellesite@fu-berlin.de}

\FUBaffiliation

\begin{abstract}
    The celebrated Lindblad equation governs the non-unitary time evolution of density operators used in the description of open quantum systems. It is usually derived from the von Neumann equation for a large system, at given physical conditions, when a small subsystem is explicitly singled out and the rest of the system acts as an environment whose degrees of freedom are traced out. In the specific case of a subsystem with variable particle number, the equilibrium density operator is given by the well-known grand canonical Gibbs state. Consequently, solving the Lindblad equation in this case should automatically yield, without any additional assumptions, the corresponding density operator in the limiting case of statistical equilibrium. Current studies of the Lindblad equation with varying particle number assume, however, the grand canonical Gibbs state \textit{a priori}: the chemical potential is externally imposed rather than derived from first principles, and hence the corresponding density operator is not obtained as a natural solution of the equation. In this work, we investigate the compatibility of grand canonical statistical mechanics with the derivation of the Lindblad equation. We propose an alternative and complementary approach to the current literature that consists in using a generalized system Hamiltonian which includes a term $\mu N$. In a previous paper, this empirically well-known term has been formally derived from the von Neumann equation for the specific case of equilibrium. Including $\mu N$ in the system Hamiltonian leads to a modified Lindblad equation which yields the grand canonical state as a natural solution, meaning that all the quantities involved are obtained from the physics of the system without any external assumptions.
\end{abstract}

\maketitle

\section{introduction}

The Lindblad equation is routinely used to analyze open quantum systems embedded in a thermodynamic reservoir \cite{Manzano2020}. The situations usually treated involve primarily the exchange of energy between system and reservoir, see e.g., Refs. \cite{Prior2010,McCauley2020} and references therein, while the exchange of matter is mostly considered for stationary currents only \cite{Manzano2014}, where the number of particles in the system can be treated as a constant. For systems with a truly variable particle number, it is well-known that the fundamental principles of quantum statistical mechanics dictate that the stationary equilibrium state is given by the grand canonical density operator. One is therefore led to the logical consequence that if the derivation of the Lindblad equation is consistent with the particular assumptions of grand canonical statistical mechanics, then one must be able to obtain the corresponding density operator from the equation in the limiting case of statistical equilibrium, without manually imposing additional physical constraints. As will be shown below, this general requirement boils down to concrete mathematical conditions which the Lindblad operator necessarily has to fulfill at equilibrium, and we will discuss their compatibility with the physical assumptions underlying the modeling of dissipative processes and the grand canonical ensemble.

Our motivation for analyzing the relationship between the Lindblad equation and the grand canonical ensemble was born out of the previous paper \cite{DelleSite2024jpa} of the second-named author, where an evolution equation for the density operator of open quantum systems close to equilibrium with a reservoir of energy and particles has been derived explicitly from first principles. Most importantly this derivation formally justifies, under the key physical approximation of negligible surface-to-volume ratio (explained below), a long-standing empirical conjecture due to Bogoliubov \cite{Bogoliubov1958}, namely that the Hamiltonian $H$ of a system in contact with a particle reservoir should be extended by the term $\mu N$, where $\mu$ is the chemical potential and $N$ the particle number operator; see, for example, Refs. \cite{Sankovich2010, Zagrebnov2001}, \cite[p. 29]{LandauLifshitz9}, \cite[p. 28]{BruusFlensberg2004}, and \cite[p. 60]{Coleman2015}. The crucial difference of the analysis conducted in Ref. \cite{DelleSite2024jpa} compared to the presentation in the aforementioned references (and similar literature in quantum many-body theory) is that the chemical potential is not introduced empirically as an external parameter, but is automatically obtained from the intrinsic physical quantities of the system. In this sense, the derivation is self-contained and does not need any empirical assumptions. With this in mind, consider the standard derivation of the Lindblad equation, as presented for example in Ref. \cite{Manzano2020}, which relies on the crucial approximation of a weak system--reservoir coupling. The important point to note is that this derivation does not make use of the assumption of negligible surface-to-volume ratio, which is physically different from the weak coupling approximation but, at the same time, mandatory for obtaining the grand canonical density operator. Thus, the question naturally arises whether a grand canonical-like variable particle number is fully compatible with the current derivation of the Lindblad equation without any further \textit{ad hoc} assumptions, and we intend to provide an answer in this work.

To avoid misconceptions, we emphasize at this point that our intention here is not to criticize current literature which treats situations involving the Lindblad equation with varying particle number in a grand canonical ensemble, such as Refs. \cite{Schaller2011, Guimaraes2016, BulnesCuetara2016, Hauptmann2019}, for example. Studies like these have delivered reliable results through approaches that are physically interesting and numerically satisfactory. In fact, we actually use them as reference for our work and aim at proposing an alternative approach that generalizes and includes the previous models on the basis of physical and mathematical consistency. For instance, in Ref. \cite{Schaller2011} the grand canonical solution of the Lindblad equation for the system in contact with a reservoir is obtained by imposing that the latter is described by a grand canonical density operator; we reach the same result without imposing such statistics \textit{a priori}. Similarly, in Ref. \cite{Guimaraes2016} the grand canonical solution is numerically obtained by empirically imposing eigenvectors of the Hamiltonian $H - \mu N$ that we have mathematically derived from first principles. In essence, therefore, our target is to obtain the grand canonical statistics without any external imposition or empirical/semi-empirical consideration, but at the same time we require complementarity to models already present in literature. In fact, our proposed extension must be fully consistent with previous models and should possibly include them; this aspect will be underlined later on for specific examples.

As a final point, we want to highlight that quantum systems in equilibrium which exchange particles with a reservoir are of high relevance in many fields of current research. Molecular physics is a typical example, where solvation properties, chemical reactions in a liquid, and similar situations would highly benefit from an open-system approach with varying particle number. The treatment of such systems via molecular simulation imposes the challenge of implementing well-founded equations, such as the Lindblad equation, in efficient codes \cite{DelleSite2017pr, DelleSite2018cpc, DelleSite2018ats}. Having applications like these in mind, the objective of the paper at hand is to analyze conceptual challenges in the derivation of the Lindblad equation for open systems with varying particle number in equilibrium, and we conclude that a possible modification of the derivation in this setting would provide a more solid foundation from the point of view of statistical mechanics. Our model consists of density operators $\rho_N$ for each $N$-particle realization of the open system and a corresponding hierarchy of equations governing the evolution of these $\rho_N$, instead of a straightforward dynamical equation for the full density operator $\rho$.

To conclude the introduction, we provide a brief outline of the paper. In \cref{sec:openSystems} the microscopic derivation of the Lindblad equation is reviewed very briefly, and a simple two-level system is discussed to highlight the difference between the exchange of excitations and the exchange of particles. In \cref{sec:GC} we introduce the grand canonical density operator, review the results of Ref. \cite{DelleSite2024jpa} mentioned above, and analyze the Lindblad equation in the grand canonical regime. In \cref{sec:revisedLindblad} we propose a modification of the derivation of the Lindblad equation in light of the foregoing analysis, leading to a revised equation directly compatible with grand-canonical statistical mechanics. We then discuss a computational protocol with which this revised Lindblad equation can be implemented in practical applications.

\section{Open Quantum Systems}\label{sec:openSystems}

In the following, the general mathematical setup for open quantum systems, which will be used throughout the paper, shall be discussed to fix notation, and the derivation of the well-known Lindblad equation will be sketched very briefly, summarizing its essence for the requirements of this work; a detailed derivation of this equation can be found in many references, e.g., Refs. \cite{Manzano2020, AlickiLendi2007, BreuerPetruccione2007, RivasHuelga2012, Vacchini2024}. We also discuss a simple example to emphasize the different physical nature of the exchange of excitations and the exchange of physical particles.

\subsection{Mathematical setup}\label{subsec:openQuantumSystems}

Consider a quantum system described by a Hilbert space $\MH$ and Hamiltonian $H$. A general (mixed) state of this system, modeled in terms of a density operator $\rho$ on $\MH$, evolves in time according to the von Neumann equation:
\begin{equation}\label{eq:vonNeumann}
    \ii \hbar \, \frac{\diff \rho(t)}{\diff t} = \bigl[H, \rho(t)\bigr] \ .
\end{equation}
The system is partitioned into a region of interest, referred to as the \emph{open system} $\SFS$, and a large \emph{thermodynamic reservoir} $\SFR$. The former is described by a Hilbert space $\MH_\SFS$ and the latter by a Hilbert space $\MH_\SFR$ such that the total space $\MH$ factorizes tensorially as \cite{Emch1968}
\begin{equation*}
    \MH = \MH_\SFS \otimes \MH_\SFR \ .
\end{equation*}
Corresponding to this factorization, the Hamiltonian $H$ is decomposed as
\begin{equation}\label{eq:Htot}
    H = H_\SFS \otimes \id_\SFR + \id_\SFS \otimes H_\SFR + \alpha H_\mathrm{int} \ ,
\end{equation}
where $H_\SFS$ is the Hamiltonian of the open system, $H_\SFR$ the Hamiltonian of the reservoir, $H_\mathrm{int}$ the Hamiltonian mediating the interaction between $\SFS$ and $\SFR$, $\id$ denotes the identity operator on the corresponding space, and $\alpha \in \R$ is the coupling constant determining the strength of the interaction. The interaction Hamiltonian $H_\mathrm{int}$ is usually decomposed as
\begin{equation}\label{eq:HintDecomp}
    H_\mathrm{int} = \sum_\ell (S_\ell \otimes R_\ell) \ ,
\end{equation}
where $S_\ell$ and $R_\ell$ are self-adjoint operators on $\MH_\SFS$ and $\MH_\SFR$, respectively; for a bounded interaction, such a decomposition is always possible according to a general result about tensor products of type I von Neumann algebras \cite[p. 185]{Takesaki1979}.

The dynamics of the total system is described in the \emph{interaction picture}, into which a general Schrödinger-picture operator $A$ can be transformed with
\begin{equation*}
    A(t) = \ee^{\ii (H_\SFS \otimes \id_\SFR + \id_\SFS \otimes H_\SFR) \, t / \hbar} \, A \, \ee^{-\ii (H_\SFS \otimes \id_\SFR + \id_\SFS \otimes H_\SFR) \, t / \hbar} \ .
\end{equation*}
In this representation, the interaction-picture density operator $\rho(t)$ evolves according to the von Neumann equation \eqref{eq:vonNeumann} with the total Hamiltonian $H$ replaced by $\alpha H_\mathrm{int}(t)$, the latter being considered in the interaction picture as well \cite{BreuerPetruccione2007}:
\begin{equation}\label{eq:vonNeumannInt}
    \frac{\diff \rho(t)}{\diff t} = - \frac{\ii \alpha}{\hbar} \, \bigl[H_\mathrm{int}(t), \rho(t)\bigr] \ .
\end{equation}

\subsection{Lindblad equation}\label{subsec:derivationLindblad}

Assuming a weak coupling between system and reservoir ($\alpha \ll 1)$, one can solve the differential equation \eqref{eq:vonNeumannInt} iteratively by integration and in this way obtain a series expansion for $\dot{\rho}(t)$. Since one is interested in the dynamical evolution of the degrees of freedom of the open system $\SFS$ only, one also applies the partial trace $\tr_\SFR$ over the reservoir $\SFR$ on this expansion; writing $\rho_\SFS(t) \ce \tr_\SFR (\rho(t))$ this yields \cite{Manzano2020}
\begin{equation*}
    \frac{\diff \rho_\SFS(t)}{\diff t} = - \frac{\ii \alpha}{\hbar} \, \tr_\SFR \bigl([H_\mathrm{int}(t), \rho(0)]\bigr) - \left(\frac{\alpha}{\hbar}\right)^{2} \int_{0}^{t} \tr_\SFR \bigl(\bigl[H_\mathrm{int}(t), [H_\mathrm{int}(s), \rho(s)]\bigr]\bigr) \diff s + \MO(\alpha^3) \ .
\end{equation*}
Since this expression still depends on the density operator $\rho(t)$ of the full system, one has to impose further assumptions to reduce the equation to the system density operator $\rho_\SFS(t)$. Namely, under the Born-Markov and rotating wave approximations, which essentially assume that the timescales of system--reservoir correlations and relaxation of the reservoir are much smaller than the typical timescale of the open system \cite{Manzano2020}, one obtains, after a series of very non-trivial manipulations, the so-called \emph{Lindblad master equation} \cite{Manzano2020, BreuerPetruccione2007, RivasHuelga2012, Vacchini2024}:
\begin{equation}\label{eq:Lindblad}
    \frac{\diff \rho_\SFS(t)}{\diff t} = - \frac{\ii}{\hbar} \, \bigl[H_\SFS + \alpha^2 H_\mathrm{ren}, \rho_\SFS(t)\bigr] + \alpha^2 \sum_j \lambda_j \left(L_j \rho_\SFS(t) L_j^\dagger - \frac{1}{2} \left\{L_j^\dagger L_j, \rho_\SFS(t)\right\}\right) \ .
\end{equation}
The first term describes the unitary (nondissipative) part of the dynamics; the role of the self-adjoint operator $H_\mathrm{ren}$ added to the system's Hamiltonian $H_\SFS$ is to renormalize the energy levels of the open system due to the interaction with the reservoir. Since these two operators commute, $H_\mathrm{ren}$ simply shifts the energy levels of $\SFS$ and is, therefore, usually referred to as the \emph{Lamb shift Hamiltonian} \cite[p. 62]{RivasHuelga2012}. The second term
\begin{equation}\label{eq:LindbladDissPart}
    \MD \bigl(\rho_\SFS(t)\bigr) \ce \alpha^2 \sum_j \lambda_j \left(L_j \rho_\SFS(t) L_j^\dagger - \frac{1}{2} \left\{L_j^\dagger L_j, \rho_\SFS(t)\right\}\right)
\end{equation}
describes the dissipative part of the dynamics in the form of a flux of energy between the system and the reservoir; this dissipation is contained in the so-called jump operators $L_j$ acting in $\MH_\SFS$ and damping rates $\lambda_j \ge 0$. All of these quantities are expressed in terms of the system and reservoir operators $S_\ell, R_\ell$ appearing in the decomposition \eqref{eq:HintDecomp} of the interaction Hamiltonian $H_\mathrm{int}$.

\begin{remark}
    Equation \eqref{eq:Lindblad} was obtained for finite-dimensional Hilbert spaces by Gorini, Kossakowski and Sudarshan \cite{Gorini1976}, and for bounded Hamiltonians on infinite-dimensional spaces by Lindblad \cite{Lindblad1976}. These authors showed in their respective cases that \eqref{eq:Lindblad} is the most general form for the generator of a quantum dynamical semigroup \cite[Theorem 5.1]{Vacchini2024}.
    
    For the sake of completeness, we mention that besides the weak-coupling derivation of \cref{eq:Lindblad} briefly discussed above, there are other, more recent methods for obtaining the Lindblad equation in different parameter regimes, e.g., coarse-graining approaches \cite{Lidar2001, Schaller2008}.
\end{remark}

\begin{remark}\label{rem:canonicalSolution}
    For later considerations, it is important to point out the following property of \cref{eq:Lindblad}: if the reservoir $\SFR$ is assumed to be in a canonical Gibbs state at inverse temperature $\beta = (k_\mathrm{B} T)^{-1} > 0$, that is, $\rho_\SFR = Z_\SFR^{-1} \, \ee^{- \beta H_\SFR}$, then it follows that the corresponding state for the open system $\SFS$ with the same inverse temperature, i.e., $\rho_\SFS = Z_\SFS^{-1} \, \ee^{- \beta H_\SFS}$, is a stationary solution of \cref{eq:Lindblad}; see, for example, Refs. \cite[Sec. 3.3.2]{BreuerPetruccione2007}, \cite[Sec. 5.2.8]{RivasHuelga2012}, and \cite[Sec. 6.3.1]{Vacchini2024}.
\end{remark}

\subsection{Exchange of excitations versus exchange of particles}\label{subsec:excitationsVSParticles}

While the Lindblad equation \eqref{eq:Lindblad} does describe the exchange of excitations or quasiparticles between system and reservoir, the exchange of actual particles is not contained in the model. To illustrate this point, we will discuss a situation often cited as describing a particle exhange and argue that it only involves the exchange of excitations; a more systematic analysis of this issue follows below in \cref{subsec:analysisCouplingOp}.

Consider the simple yet ubiquitous setting of a two-level system $\SFS$ with Hilbert space $\MH_\SFS = \C^2$, standard basis $\set{\ket{1}, \ket{0}}$, and Hamiltonian
\begin{equation*}
    H_\SFS = \frac{\hbar \omega_0}{2} \bigl(\ketbra{1}{1} - \ketbra{0}{0}\bigr) \ , \quad \omega_0 > 0 \ ,
\end{equation*}
coupled to a bosonic reservoir $\SFR$. If this reservoir is in a canonical Gibbs state at inverse temperature $\beta$, one can show that the dissipative part \eqref{eq:LindbladDissPart} of the Lindblad equation takes the following form \cite[Sec. 3.4.2]{BreuerPetruccione2007}, \cite[Sec. 5.2.5]{RivasHuelga2012}:
\begin{align}\label{eq:twolevelDissipator}
     \begin{split}
        \MD \bigl(\rho_\SFS(t)\bigr) &= \gamma_0 (N + 1) \bigl(\sigma_- \rho_\SFS(t) \sigma_+ - \frac{1}{2} \, \{\sigma_+ \sigma_-, \rho_\SFS(t)\}\bigr) \\
        &\quad + \gamma_0 N \bigl(\sigma_+ \rho_\SFS(t) \sigma_- - \frac{1}{2} \, \{\sigma_- \sigma_+, \rho_\SFS(t)\}\bigr) \ .
    \end{split}
\end{align}
$\gamma_0 > 0$ is a physical constant (depending on $\omega_0$), $N = \bigl(\ee^{\beta \hbar \omega_0} - 1\bigr)^{-1}$ is the value of the Bose-Einstein distribution at $\omega_0$, that is, the mean number of bosons with frequency $\omega_0$ in the thermal state of the reservoir, and the matrices $\sigma_{\pm}$ are the \enquote{ladder operators} for the two-level system, that is, $\sigma_+ = \ketbra{1}{0}$ and $\sigma_- = \ketbra{0}{1}$. The two terms above can be interpreted as the transfer of an \emph{excitation} (that is, a quasiparticle) from the system to the reservoir at rate $\gamma_0 (N + 1)$ and from the reservoir to the system at rate $\gamma_0 N$, respectively \cite[p. 148]{BreuerPetruccione2007}, \cite[p. 320]{Vacchini2024}.

This exchange does not constitute an exchange of actual particles in the grand canonical sense; the only process which is modeled by \cref{eq:twolevelDissipator} is a jump of a single particle between the two states $\ket{0}, \ket{1}$, induced by the environment of bosonic particles. In fact, for an actual exchange of particles, one would need (i) a chemical potential provided by the physics of the system which is, however, not modelded in this jump process, and (ii) a change of the intrinsic physical constitution of the system $\SFS$ due to the addition or removal of degrees of freedom. The latter should be reflected by the system's Hilbert space $\MH_\SFS$ which stays the same in the above example though (as it still models only a single two-level degree of freedom), but would need to change if the number of particles contained in $\SFS$ changes.

Incorporating such processes involving the exchange of actual particles in the Lindblad equation is the novel aspect of our paper which we will discuss in detail below.

\begin{remark}
    A step forward compared to this approximation of constant particle number in the subsystem can already be found in the seminal work of Emch and Sewell \cite{Emch1968}, where they suggest a coupling term composed of creation and annihilation operators for particles in the system. Despite the fact that their derivation is very general and even goes beyond the approximation of a Markovian reservoir, they remain rather unspecific regarding the form that such an operator should have. In fact, a solution is not suggested even in the simplified case of equilibrium.
\end{remark}

\section{Grand canonical ensemble}\label{sec:GC}

To carry out a more systematic analysis of the Lindblad equation for systems with variable particle number, going beyond the example discussed above, we shall introduce the grand canonical formalism in this section. We will also briefly report on the recent first-principle derivation of an effective Hamiltonian for open quantum systems with varying particle number, mentioned in the introduction. We shall then return to the Lindblad equation and argue that the grand canonical density operator cannot be obtained from it without imposing additional assumptions.

\subsection{Fock space}

Consider the general setup discussed in \cref{subsec:openQuantumSystems}, and suppose that the open quantum system $\SFS$, consisting of identical particles, exchanges energy and matter with the reservoir $\SFR$. In this case, the state space $\MH_\SFS$ of $\SFS$ is naturally given by the Fock space
\begin{equation*}
    \MF \ce \bigoplus_{N=0}^{\infty} \MH^{\otimes N} \ ,
\end{equation*}
where $\MH$ is the Hilbert space of a single particle in $\SFS$ and $\MH^{\otimes N} \ce \bigotimes^N \MH$ (with the convention $\MH^{\otimes 0} \equiv \C$) the $N$-fold tensor product of $\MH$, that is, the state space of a realization of the open system comprising $N$ particles (referred to as an \emph{$N$-realization} in the following). Elements of $\MF$ are sequences $(\Phi_N)_{N \in \N_0}$ of vectors $\Phi_N \in \MH^{\otimes N}$ such that $\sum_{N=0}^{\infty} \norm{\Phi_N}_N^2 < + \infty$, with $\norm{\,\cdot\,}_N$ being the canonical norm on the tensor product space $\MH^{\otimes N}$.

On the Fock space $\MF$, one can define the total particle number operator $\MSN$ as the direct sum operator
\begin{equation}\label{eq:FockNumberOp}
    \MSN = \bigoplus_{N=0}^{\infty} \bigl(N \id_{\MH^{\otimes N}}\bigr) \ ,
\end{equation}
which means that $\MSN$ acts on an element $(\Phi_N)_{N \in \N_0} \in \MF$ according to $\MSN (\Phi_N)_{N \in \N_0} = (N \Phi_N)_{N \in \N_0}$. Similarly, if $H_N$ denotes the Hamiltonian of an $N$-particle realization of the system $\SFS$, acting in the tensor product space $\MH^{\otimes N}$, then its extension $\MSH$ to Fock space $\MF$, usually called \enquote{second quantization} of the single-particle operator, is given by \cite{BratteliRobinson1981}
\begin{equation}\label{eq:FockHamiltonian}
    \MSH = \bigoplus_{N=0}^{\infty} H_N \ .
\end{equation}
Two important mathematical properties, which will be used below, readily follow from the definitions of the operators \eqref{eq:FockNumberOp} and \eqref{eq:FockHamiltonian}: first, these two operators commute, $\bigl[\MSH, \MSN] = 0$, and second, restricted to the subspace $\MH^{\otimes N}$ they act like $N \id_{\MH^{\otimes N}}$, respectively, $H_N$.

\subsection{Grand canonical Gibbs state}

The equilibrium state of the open system $\SFS$, exchanging energy and particles with the reservoir $\SFR$, at inverse temperature $\beta$ is given by the well-known grand canonical density operator \cite{Huang1991, Schwabl2006}
\begin{equation}\label{eq:GCdensityMatrix}
    \rho_\mathrm{GC} \ce \frac{1}{Q_\mathrm{GC}} \, \ee^{- \beta (\MSH - \mu \MSN)} \ ,
\end{equation}
where $\mu \in \R$ is the chemical potential, that is, the rate at which the exchange of particles with the reservoir occurs, and $Q_\mathrm{GC}$ is a normalization factor given by
\begin{equation*}
    Q_\mathrm{GC} \ce \tr_\MF \Bigl(\ee^{- \beta (\MSH - \mu \MSN)}\Bigr) = \sum_{N=0}^{\infty} \tr_{\MH^{\otimes N}} \bigl(\ee^{- \beta (H_N - \mu N \id)}\bigr) \equiv \sum_{N=0}^{\infty} Q_N \ .
\end{equation*}
Using the (nonnormalized) trace-class operator
\begin{equation}\label{eq:rhoN}
    \rho_{\mathrm{GC}, N} \ce \frac{1}{Q_\mathrm{GC}} \, \ee^{- \beta (H_N - \mu N \id_{\MH^{\otimes N}})} \ ,
\end{equation}
which acts in the $N$-particle Hilbert space $\MH^{\otimes N}$, one may rewrite the grand canonical Gibbs state \eqref{eq:GCdensityMatrix} as the direct sum operator
\begin{equation}\label{eq:GCrhoN}
    \rho_\mathrm{GC} = \bigoplus_{N=0}^{\infty} \rho_{\mathrm{GC}, N} = \frac{1}{\sum_{N=0}^{\infty} Q_N} \, \bigoplus_{N=0}^{\infty} \ee^{- \beta (H_N - \mu N \id_{\MH^{\otimes N}})} \ .
\end{equation}
This relation shows that instead of specifying the full Fock space density operator $\rho_\mathrm{GC}$, one may work with the hierarchy of $N$-particle operators $\rho_{\mathrm{GC}, N}$ and the corresponding normalization factors $Q_N$; this observation will be crucial later on.

\subsection{First-principle derivation of the grand canonical density operator}\label{subsec:Heff}

In Ref. \cite{DelleSite2024jpa}, the effective Hamiltonian $H_N - \mu N \id$ for open quantum systems exchanging energy and particles with their environment was derived from first principles. This derivation provides a theoretical justification for a long-standing empirical conjecture used frequently in quantum many-body theory as well as for a numerical algorithm used in path integral molecular dynamics simulations of systems in contact with a particle reservoir. Employing this effective Hamiltonian, the recent follow-up study \cite{Reible2025} demonstrated the utility of a variable particle number as a tool to control the accessible spectrum of a quantum system, thus strengthening the utility of such an effective Hamiltonian by opening up new possibilities for the analysis of quantum systems.

The derivation proposed in Ref. \cite{DelleSite2024jpa} follows a statistical mechanics approach rather than the purely dynamical viewpoint taken in the derivation of the Lindblad equation. In essence, the effective Hamiltonian is derived by separating a large system into a thermodynamic reservoir $\SFR$ in which a system of interest $\SFS$ is embedded, and then tracing out the degrees of freedom of $\SFR$ in the von Neumann equation \eqref{eq:vonNeumann} for a generic $N$-realization of $\SFS$, using the approximation of negligible surface-to-volume ratio. To this end, the Hamiltonian $H$ of the total system is first partitioned as in \cref{eq:Htot}. Contrary to the derivation of the Lindblad equation, however, it is then assumed that $\norm{\alpha H_\mathrm{int}} \ll \norm{H_\SFS}$, hence
\begin{equation}\label{eq:surfaceToVolumeRatio}
    H \approx H_\SFS \otimes \id_\SFR + \id_\SFS \otimes H_\SFR \ .
\end{equation}
This is the approximation of \emph{negligible surface-to-volume ratio} known from statistical mechanics \cite{Huang1991, Schwabl2006}, and it represents the key difference compared to the derivation of the Lindblad equation. Indeed, this approximation is not a subcase of the weak coupling approximation: the direct interaction between system and reservoir can be locally strong, while the statistical weight of the interface energy is small compared to the total internal energy, thus defining a minimal volume for the subsystem of interest.

Next, assume (i) that the total system contains $M$ particles and is partitioned such that $\SFS$ contains $N$ (instantaneous) particles, as done similarly for classical systems in the Liouville equation \cite{DelleSite2020, Klein2022}; we shall denote the system Hamiltonian $H_\SFS$ by $H_N$ and the bath Hamiltonian $H_\SFR$ by $H_{M - N}$ to emphasize the dependence on the partition. Assume (ii) that the reservoir is much larger than the system of interest, i.e., $N \ll M$, and furthermore that $\SFR$ is in a canonical Gibbs state $\rho_R \sim \ee^{- \beta H_{M-N}}$, independently from its interaction with $\SFS$ (which is reasonable given that $\SFR$ is much larger than $\SFS$). Writing $\rho_N(t) \ce \tr_\SFR (\rho(t))$, the von Neumann equation \eqref{eq:vonNeumann} turns into the following $N$-hierarchy of equations for the density operators $\rho_N(t)$:
\begin{equation*}
    \ii \hbar \, \frac{\diff \rho_N(t)}{\diff t} = \tr_\SFR \bigl([H, \rho(t)]\bigr) \ .
\end{equation*}

Using the surface-to-volume ratio approximation \eqref{eq:surfaceToVolumeRatio}, one can evaluate the partial trace on the right-hand side of this equation to obtain
\begin{equation*}
    \tr_\SFR \bigl([H, \rho(t)]\bigr) = [H_N, \rho_N(t)] + \tr \bigl(H_{M - N} \rho_\SFR(t)\bigr) \, \rho_N(t) - \rho_N(t) \tr \bigl(\rho_\SFR(t) H_{M - N}\bigr) \ .
\end{equation*}
While the last two terms, in principle, cancel each other out, it is fruitful to evaluate them explicitly as follows. The expressions $\tr (H_{M - N} \rho_\SFR) = \tr (\rho_\SFR H_{M - N}) \ec \braket{E_\SFR (M - N)}$ correspond to the average energy of the reservoir $\SFR$. Now, crucially, one can expand the unknown function $N \mapsto \braket{E_\SFR (M - N)}$ in a Taylor series in powers of $N$, given that the bath $\SFR$ is much larger than the system $\SFS$, i.e., $M - N \approx M$ for $N \ll M$. To first order in $N$, one therefore has
\begin{equation}\label{eq:energySeriesExpansion}
    \braket[\big]{E_\SFR (M - N)} \approx \braket[\big]{E_\SFR (M)} + \frac{\partial}{\partial N} \braket[\big]{E_\SFR (M - N)} \bigg\vert_{N \ll M} \, N \ .
\end{equation}
The first term $\braket{E_\SFR (M)}$ is a constant that determines the zero of the energy scale and can hence, without loss of generality, be chosen equal to zero; note that it plays an analogous role as the Lamb shift term in the Lindblad equation. More interestingly, the expression
\begin{equation*}
  \frac{\partial}{\partial N} \braket[\big]{E_\SFR (M - N)} \bigg\vert_{N \ll M} \equiv - \mu
\end{equation*}
corresponds, by definition, to the chemical potential $\mu$ at constant entropy and volume of the reservoir \cite[p. 71]{LandauLifshitz5}. Since we consider $\SFS$ and $\SFR$ in equilibrium, $\mu$ is automatically the chemical potential of the system $\SFS$ as well. This point is truly crucial for the subject discussed in this paper: it highlights that, given the total system, the explicit derivation of the chemical potential is naturally obtained from the von Neumann equation according to the general physical approximations made at the beginning. The key point is, in particular, that $\mu$ is not imposed {\it a posteriori} (as done, for example, in Refs. \cite{Schaller2011, Guimaraes2016, BulnesCuetara2016, Hauptmann2019}), which implies that the ensemble consistency and the corresponding thermodynamics in Ref. \cite{DelleSite2024jpa} naturally emerge from the physics of the system, as they should.

Finally, observe that one may lift the expression \eqref{eq:energySeriesExpansion} to an operator identity on the $N$-particle subspace $\MH^{\otimes N}$:
\begin{equation*}
    \braket[\big]{E_\SFR (M - N)} \approx - \mu N \id_{\MH^{\otimes N}} \ .
\end{equation*}
This is a reasonable assumption since according to Eq. \eqref{eq:FockNumberOp}, the number operator $\MSN$ will simply count the number of particles of the $N$-realization of $\SFS$ and thus give back the value $N$. With this, the resulting equation for the density operator $\rho_N(t)$ of the open system $\SFS$ near equilibrium becomes
\begin{equation}\label{eq:vonNeumannGC}
    \ii \hbar \, \frac{\diff \rho_{N}(t)}{\diff t} = \bigl[H_N - \mu N \id_{\MH^{\otimes N}}, \rho_{N}(t)\bigr] \ .
\end{equation}
This equation suggests that an $N$-realization of the system $\SFS$ is described by the effective Hamiltonian
\begin{equation*}
    H_N^\mathrm{eff} \ce H_N - \mu N \id_{\MH^{\otimes N} \ ,}
\end{equation*}
with a clear definition of the chemical potential $\mu$ which emerges from the physical quantities involved in the derivation. In particular, in the case of stationary equilibrium, this equation automatically delivers the grand canonical density operator in the form of \cref{eq:rhoN}.

\begin{remark}
    As mentioned above, the basic tool used in the foregoing derivation is the well-known surface-to-volume ratio approximation, commonly used in physics when a system is divided into two subsystems such that the cross-interactions between them are negligible compared to the internal interactions of each of them. It is important to note that one can even rigorously estimate whether the correction due to the separation is negligible or not, see Refs. \cite{DelleSite2017jsm,Reible2022,Reible2023,DelleSite2024pra,DelleSite2024molphys}. Assuming that the separation is justified, the problem simplifies significantly while still capturing the relevant physics of the situation. For the specific case of an open system which is large enough to be statistically well-defined, embedded in a much larger reservoir whose macroscopic properties are not affected by the open system, this approximation is certainly well-founded.
\end{remark}

\subsection{Lindblad coupling operator in the grand canonical regime}\label{subsec:analysisCouplingOp}

Having introduced the grand canonical Gibbs state \eqref{eq:GCdensityMatrix} in the natural language of Fock space and outlined how it emerges in a first-principle derivation, we shall now return to the Lindblad equation. In the limiting case of stationary equilibrium, it holds that $\dot{\rho}_\SFS = 0$, and hence \cref{eq:Lindblad} reduces to
\begin{equation*}
    -\frac{\ii}{\hbar} \, \bigl[H_\SFS, \rho_\SFS\bigr] + \alpha^2 \sum_j \lambda_j \left(L_j \rho_\SFS L_j^\dagger - \frac{1}{2} \left\{L_j^\dagger L_j, \rho_\SFS\right\}\right) = 0 \ ;
\end{equation*}
for simplicity, we have neglected the renormalizing shift $H_\mathrm{ren}$ in the system's energy. If one now considers the specific case of an open system $\SFS$ embedded in a large reservoir with which it exchanges energy and particles, then the density operator $\rho_\SFS$ is given by $\rho_\mathrm{GC}$ from \cref{eq:GCdensityMatrix}, and the Hamiltonian $H_\SFS$ is the second quantization operator $\MSH$ from \cref{eq:FockHamiltonian}. Inserting this into the previous identity, it follows that
\begin{equation*}
    \frac{\ii}{\hbar} \, \frac{1}{Q_\mathrm{GC}} \bigl[\MSH, \ee^{- \beta (\MSH - \mu \MSN)}\bigr] = \frac{\alpha^2}{Q_\mathrm{GC}} \, \sum_j \lambda_j \left(L_j \, \ee^{- \beta (\MSH - \mu \MSN)} L_j^\dagger - \frac{1}{2} \left\{L_j^\dagger L_j, \ee^{- \beta (\MSH - \mu \MSN)}\right\}\right) \ .
\end{equation*}
Observe that the left-hand side vanishes since $[\MSH, \ee^{- \beta (\MSH - \mu \MSN)}] = 0$, which follows from $[\MSH, \MSH - \mu \MSN] = 0$. Therefore, the condition on the damping rates $\lambda_j$ and the jump operators $L_j, L_j^\dagger$ for the equilibrium case is
\begin{equation}\label{eq:equilibriumCondition}
    \sum_j \lambda_j \left(L_j \, \ee^{- \beta (\MSH - \mu \MSN)} L_j^\dagger - \frac{1}{2} \left\{L_j^\dagger L_j, \ee^{- \beta (\MSH - \mu \MSN)}\right\}\right) = 0 \ .
\end{equation}
Three possibilities arise to satisfy \cref{eq:equilibriumCondition}:
\begin{enumerate}[wide=\parindent, label=(\Alph*)]
    \item \label{enu:cond1} The operators $L_j$ and $L_j^\dagger$ are such that for all $j$:
    \begin{equation*}
        L_j^\dagger L_j = L_j L_j^\dagger \ , \quad \bigl[L_j, \ee^{- \beta (\MSH - \mu \MSN)}\bigr] = 0 \ , \quad \text{and} \quad \bigl[L_j^\dagger, \ee^{- \beta (\MSH - \mu \MSN)}\bigr] = 0 \ .
    \end{equation*}
    That is, the operators $L_j$ are normal, and $\ee^{- \beta (\MSH - \mu \MSN)}$ commutes with $L_j$ and $L_j^\dagger$. (Instead of the latter, one may also require that $L_j$ and $L_j^\dagger$ commute with the operator $\MSH - \mu \MSN$.)

    \item \label{enu:cond2} The involved system--reservoir processes balance each other in the sum, that is, there exist numbers $r, s \in \N$ such that $r + s$ equals the total number of processes and such that, after possibly relabeling, one has
    \begin{equation*}
        \begin{gathered}
            \sum_{j=1}^{r} \lambda_j \left(L_j \, \ee^{- \beta (\MSH - \mu \MSN)} L_j^\dagger - \frac{1}{2} \left\{L_j^\dagger L_j, \ee^{- \beta (\MSH - \mu \MSN)}\right\}\right) \\
            = - \sum_{j=r+1}^{s} \lambda_j \left(L_j \, \ee^{- \beta (\MSH - \mu \MSN)} L_j^\dagger - \frac{1}{2} \left\{L_j^\dagger L_j, \ee^{- \beta (\MSH - \mu \MSN)}\right\}\right) \ .
        \end{gathered}
    \end{equation*}
    This is to say that one can specifically build (e.g., in a numerical scheme) processes that, by design, balance each other.

    \item \label{enu:cond3} The sum of dissipative processes associated to the jump operators $L_j, L_j^\dagger$ and damping rates $\lambda_j$ is forced into an effective total nondissipative process, i.e.,
    \begin{equation}\label{eq:effgc}
        \begin{gathered}
            \sum_j \lambda_j \left(L_j \, \ee^{- \beta (\MSH - \mu \MSN)} L_j^\dagger - \frac{1}{2} \left\{L_j^\dagger L_j, \ee^{- \beta (\MSH - \mu \MSN)}\right\}\right) \\
            = \frac{\ii}{\hbar} \, \bigl[f(\mu \MSN), \ee^{- \beta (\MSH - \mu \MSN)}\bigr] \ .
        \end{gathered}
    \end{equation}
    Here, $f = f(\mu \MSN)$ is some operator-valued function of the grand canonical system--reservoir coupling. In particular, $f$ is not a function of the open system Hamiltonian $\MSH$, since the latter does not carry any explicit information about the coupling to the reservoir.
\end{enumerate}

The conditions \ref{enu:cond1} and \ref{enu:cond2}, on the one hand, would restrict the specific modeling of the system--reservoir coupling to a very limited amount of physical possibilities, if there are any at all.  While the specific choice of operators would be matching the limiting case of equilibrium, it would not necessarily describe situations out of equilibrium. Condition \ref{enu:cond2}, in particular, represents a specific form of imposed detailed balance; this is a necessary condition for a grand canonical ensemble, but it is not sufficient. In fact, in a grand canonical ensemble the particle number needs to fluctuate according to the natural thermodynamics of the system; that is, for any particle entering or leaving the subsystem, an automatic sampling is required which searches for the minimum of the free energy \cite{Widom1963}. This process is not likely to be written in a treatable form of operators without assuming, \emph{a priori}, the grand canonical eigenstates, and probably even if one makes this assumption, the form of the specific operators may be extremely complex and thus intractable in, e.g., numerical schemes.

On the other hand, condition \ref{enu:cond3} leads, in our view, to a situation that contradicts the standard derivation of the grand canonical density operator in statistical mechanics. Indeed, reinserting \cref{eq:effgc} into the Lindblad equation \eqref{eq:Lindblad} results in
\begin{equation*}
    \frac{\diff \rho_\SFS(t)}{\diff t} = - \frac{\ii}{\hbar} \bigl[\MSH, \rho_\SFS(t)\bigr] + \frac{\ii}{\hbar} \, \bigl[f(\mu \MSN), \rho_\SFS(t)\bigr] \ ,
\end{equation*}
thus implying an effective Hamiltonian of equilibrium of the form $\MSH - f(\mu \MSN)$. At first order of the operator-valued function $f(\mu \MSN)$, this gives $\MSH - \mu \MSN$ which is formally equivalent to the effective Hamiltonian that was derived in Ref. \cite{DelleSite2024jpa} under the approximation of negligible surface-to-volume ratio, meaning that $H_\mathrm{int}$ from the decomposition \eqref{eq:Htot} must be neglected overall. Since the derivation of the Lindblad equation crucially relies on the system--reservoir interaction Hamiltonian $H_\mathrm{int}$, this would imply that the same effective Hamiltonian $\MSH - \mu \MSN$ would follow from condition \ref{enu:cond3} without the aforementioned approximation, thereby contradicting the statistical mechanics of the grand canonical ensemble.

Based on this discussion, we conclude that none of the three conditions \ref{enu:cond1}, \ref{enu:cond2}, or \ref{enu:cond3} can be achieved without great loss of generality, hence the grand canonical density operator is not a natural solution of the Lindblad equation, without additional assumptions.

\begin{remark}\label{rem:chemostat}
    It must be noted that for a system with fixed particle number, the external coupling that drives the system to thermodynamic equilibrium, e.g., a thermostat, can be switched off once the thermalization is reached; thus, the equivalent of \cref{eq:equilibriumCondition} for the canonical Gibbs state is automatically achieved in equilibrium by $\lambda_j \to 0$. This is \emph{not} true for a chemostat, however, as the transfer of particles and energy from and to the reservoir requires a continuous exchange. Moreover, the two situations also differ with respect to the definition of the thermodynamic state point. In the canonical case, the only thermodynamic variables are particle density and temperature, which are fixed by the external observer. Instead, in a grand canonical set up one needs, in addition, knowledge of the chemical potential which cannot be arbitrarily fixed by an observer, but must be automatically derived through the physical quantities of the system; this point will be very important later on for the model we propose.
\end{remark}

\begin{remark}
    Regarding alternative approaches to grand canonical systems within the framework of the Lindblad equation hinted at in the introduction, we mention first the important results of Ref. \cite{Schaller2011}. In this paper, the author showed, among other things, that \emph{assuming} the reservoir $\SFR$ to be in a grand canonical Gibbs state $\rho_\SFR \sim \ee^{- \beta (H_\SFR - \mu N_\SFR)}$, it follows that the corresponding density operator for the open system at the same inverse temperature $\beta$ and chemical potential $\mu$, $\rho_\SFS \sim \ee^{- \beta (H_\SFS - \mu N_\SFS)}$, is a stationary solution of the Lindblad equation \eqref{eq:Lindblad}. This analysis is definitely nontrivial and interesting as it generalizes the property mentioned in \cref{rem:canonicalSolution} to grand canonical situations. However, the chemical potential of the reservoir has to be imposed manually and is not derived from physical quantities of the total system. As discussed above, the chemical potential cannot be fixed \emph{a priori} like the temperature, and hence, in our view, it does not suffice to generalize \cref{rem:canonicalSolution} to obtain grand canonical consistency. Instead, the chemical potential has to be provided from first principles when solving the Lindblad equation, and this is exactly what we desire to achieve in this paper.
    
    Next, we comment on the interesting approach of Ref. \cite{Guimaraes2016} which treats quantum chains with Lindblad baths. In this study, the coupling coefficients of the dissipative part \eqref{eq:LindbladDissPart} of the Lindblad equation are assumed to have a grand canonical form of number occupation, with the additional imposition of a chemical potential (similar to the approach of Ref. \cite{BulnesCuetara2016}). In particular, the authors of Ref. \cite{Guimaraes2016} manually impose the eigenstates of the operator $H_N - \mu N \id_S$ to conclude about the detailed balance of the process. While this approach is certainly very interesting, it also relies on an \emph{a posteriori} imposition of the system--reservoir coupling term into the Lindblad equation that does not naturally emerge from tracing out the degrees of freedom of the reservoir. In relation to our model, as discussed below, the employed Hamiltonian reminds of the result of Ref. \cite{DelleSite2024jpa}, thus implicitly suggesting that one should separate the equilibrium state from the nonequilibrium state and add the nonequilibrium operator as a perturbation over the equilibrium state.
\end{remark}

\section{Revised Lindblad equation for grand canonical consistency}\label{sec:revisedLindblad}

After a quick summary of the foregoing discussion, we shall present our proposal for overcoming the difficulties discussed in the previous subsection. This will lead us to a modified Lindblad equation in the form of an $N$-hierarchy of equations for the $N$-particle density operators. We will also discuss a possible computational protocol associated with this hierarchy at the end of the section.

\subsection{Summary of the previous discussion}

The essential difference between the derivation of the Lindblad equation and the effective Hamiltonian lies in the partitioning of the total system and the decomposition of its Hamiltonian. In the former, the operator that drives the time evolution is the part $\alpha H_\mathrm{int}$ of the total Hamiltonian that mediates the interaction between the open system and the reservoir, cf. \cref{eq:vonNeumannInt}. Contrarily, in the latter the key approximation is that of a negligible surface-to-volume ratio, implying that the part of the total Hamiltonian carrying the system--reservoir interaction can be dropped. In equilibrium statistical mechanics, this is not a drastic approximation: the exchange of particles between the open system and the environment happens according to the balance of free energy (that is, equal chemical potential); if the environment is large enough, then its free energy will not be affected by the change in the number of particles due to the exchange between system and reservoir, and thus one can effectively take the chemical potential of the total system as the chemical potential of the reservoir. In turn, to reach equilibrium, the open system must adjust its free energy according to the free energy of the reservoir. Hence, the interactions between the open system and the reservoir, despite the fact that they are not explicitly considered, take place statistically and are, therefore, implicitly included in the free energy of the reservoir and in the corresponding \enquote{response} of the open system driven by the reservoir-imposed chemical potential. In our view, this is the key aspect that does not allow the Lindblad approach to be straightforwardly compatible with the statistical mechanics derivation of the grand canonical density operator. As argued in \cref{subsec:analysisCouplingOp}, the derivation of the Lindblad equation in its current form does not support the notion of a chemical potential automatically derived from the free energy of the environment, but it can only be empirically imposed in the final form of the equation.

\subsection{Revising the derivation of the Lindblad equation}

In the following, we propose a statistical mechanics approach to address this problem, whose key feature is to consider every $N$-realization of our system and its corresponding statistical weight. This is formally analogous to the standard approach used in classical statistical mechanics for the grand canonical ensemble, where the subsystem is characterized by an $N$-generic realization of particles \cite{DelleSite2020, Klein2022}. In fact, a natural way to proceed is to take into account the results of Ref. \cite{DelleSite2024jpa} and employ the effective Hamiltonian $H_N - \mu N \id_S$, which comes with a first-principle derivation of the chemical potential, as the Hamiltonian of the open system $\SFS$ in the Lindblad equation, so that, in the limit of vanishing dissipative processes (that is, in stationary equilibrium), the Lindblad equation assumes the form of \cref{eq:vonNeumannGC} of the von Neumann equation for an open system with grand canonical statistics.

To implement this idea, one can approximate the interaction Hamiltonian $\alpha H_\mathrm{int}$ between the open system and the environment as the sum of two terms: (i) a statistically conservative part $\mu N \id_\SFS$ which commutes with the system Hamiltonian $H_N$ and thus conserves equilibrium, and (ii) a dissipative part $\wt{\alpha} H_\mathrm{diss}$ that controls the convergence toward an equilibrium or nonequilibrium steady state. Thus, we set
\begin{equation}\label{eq:revisedHint}
    \alpha H_\mathrm{int} = - \mu N \id_\SFS \otimes \id_\SFR + \wt{\alpha} H_\mathrm{diss} \ ,
\end{equation}
with $H_\mathrm{diss}$ acting in the tensor product space $\MH_\SFS \otimes \MH_\SFR$ and $\wt{\alpha} \in \R$ controlling the dissipative coupling. The difference of \eqref{eq:revisedHint} compared to the usual derivation (see \cref{subsec:derivationLindblad}) is that it contains a statistically conservative part instead of only a dissipative part. In this context, \enquote{statistically conservative} is understood in the sense that the operator $\mu N \id_\SFS$ allows, in average, an exchange in equilibrium, where equilibrium means that the system is characterized by a mean energy and a mean number of particles according to the statistical ensemble average. Note also that the additional term $\mu N \id_\SFS$ mediating the particle flux in the interaction Hamiltonian \eqref{eq:revisedHint} cannot be decomposed as an operator in $\MH_\SFS \otimes \MH_\SFR$ like in \cref{eq:HintDecomp}, other than the trivial decomposition $- \mu N \id_\SFS \otimes \id_\SFR$ that was used in \cref{eq:revisedHint}. This is due to the fact that, as shown in \cref{subsec:Heff}, the chemical potential $\mu$, when derived from the partitioning of the large system, depends on the energy of the reservoir $\SFR$ and thus already contains information about the interaction with $\SFR$ which makes a further tensorial decomposition superfluous \footnote{In addition, from a mathematical point of view the choice $- \mu N \id_\SFS \otimes \id_\SFR$ is also very natural because the mapping $\BO(\MH_\SFS) \to \BO(\MH_\SFS \otimes \MH_\SFR)$, $T \mapsto T \otimes \id_\SFR$, extending a bounded operator on $\MH_\SFS$ to one on $\MH_\SFS \otimes \MH_\SFR$, is the dual, with respect to the Hilbert-Schmidt inner product, of the partial trace $\tr_\SFR : \BO_1(\MH_\SFS \otimes \MH_\SFR) \to \BO_1(\MH_\SFS)$, $S \otimes T \mapsto S \tr (T)$, mapping a trace-class operator from $\MH_\SFS \otimes \MH_\SFR$ to one on $\MH_\SFS$; see, for example, Ref. \cite[p. 122]{Petz2008}.}; this reinforces that the choice \eqref{eq:revisedHint} for the interaction is well-founded.

Now, with the model \eqref{eq:revisedHint} for the interaction Hamiltonian, the term $\mu N \id_\SFS$ can automatically be included in the effective Hamiltonian of the open system, while $H_\mathrm{diss}$ is used for the explicit coupling between the system and reservoir in any kind of nonequilibrium, dissipative process, as in the standard derivation of the Lindblad equation. For the latter, this means that instead of \cref{eq:Htot} one employs the following total Hamiltonian:
\begin{equation*}
    H = \bigl(H_N - \mu N \id_\SFS\bigr) \otimes \id_\SFR + \id_\SFS \otimes H_\SFR + \wt{\alpha} H_\mathrm{diss} \ .
\end{equation*}
The dissipative Hamiltonian $H_\mathrm{diss}$ can be decomposed as in \cref{eq:HintDecomp} since it still acts in the tensor product space. With this, the starting point for the derivation of the modified Lindblad equation is \cref{eq:vonNeumannInt} with $H_\mathrm{int}(t)$ replaced by $H_\mathrm{diss}(t)$. One can then perform the same weak-coupling expansion with respect to the parameter $\wt{\alpha}$, where all the approximations now pertain to $H_\mathrm{diss}$. The resulting equation for an $N$-realization of $\SFS$ thus reads
\begin{align}\label{eq:modifiedLindblad}
    \begin{split}
        \frac{\diff \rho_N(t)}{\diff t} &= - \frac{\ii}{\hbar} \, \bigl[H_N - \mu N \id_\SFS + \wt{\alpha}^2 H_\mathrm{ren}, \rho_N(t)\bigr] \\
        &\quad + \wt{\alpha}^2 \sum_j \lambda_j \left(L_j \rho_N(t) L_j^\dagger - \frac{1}{2} \left\{L_j^\dagger L_j, \rho_N(t)\right\}\right) \ .
    \end{split}
\end{align}
Here, the jump operators $L_j$ and the damping rates $\lambda_j$ are computed with respect to the tensorial decomposition of the dissipative interaction $H_\mathrm{diss}$. Note, in particular, that in the limit $\wt{\alpha} \to 0$ of vanishing dissipative interaction (meaning that the environment does not create any situation far from equilibrium for the open system), \cref{eq:modifiedLindblad} reduces to \cref{eq:vonNeumannGC}, and hence a stationary equilibrium state with the standard grand canonical effective Hamiltonian $H_N - \mu N \id_\SFS$ is automatically obtained, where the chemical potential is not imposed externally but derived as in \cref{subsec:Heff}.

\begin{remark}
    Observe that in this derivation the Lamb shift Hamiltonian $H_\mathrm{ren}$ corresponds to the term $\braket{E_\SFR(M)}$ in \cref{eq:energySeriesExpansion}. The term $\mu N \id_\SFS$, while in this representation also being a constant shift, depends on the variable $N$ and contains, through the chemical potential $\mu$, the statistical interaction with the environment.
    
    In principle, one could consider the modified Lindblad equation also in the full Fock space, where the variable $N$ corresponds to the number of occupied states, the $N$-particle density operator is replaced by the full Fock state $\rho_\SFS$, and the system Hamiltonian becomes $\MSH - \mu \MSN$. Aside from this, the form of the modified Lindblad equation for the full density operator $\rho$ remains unchanged compared to \cref{eq:modifiedLindblad}.

    Furthermore, we point out that on a purely empirical basis, where the term $\mu \MSN$ is simply added to the system Hamiltonian $\MSH$ for similarity to the grand canonical Boltzmann factor, a form of the Lindblad equation similar to our proposal \eqref{eq:modifiedLindblad} has already been written in the preprint \cite{Hauptmann2019}; this highlights that the need for extending the equation was already present in the community. In light of this, our novel contribution consists in introducing our results of Ref. \cite{DelleSite2024jpa} into the derivation of the Lindblad equation and completing it for the case of varying $N$ in a grand canonical fashion. Crucially, $\mu$ is then exactly defined by quantities of the system and it can be calculated explicitly:
    \begin{equation*}
        \mu = - \frac{\partial}{\partial N} \braket[\big]{E_\SFR (M - N)} \bigg\vert_{N \ll M} \ .
    \end{equation*}
    That is, $\mu$ is no longer a parameter that is defined outside the derivation of the equation, but it naturally emerges within the derivation as well.

    Finally, note that the algorithm used in molecular dynamics simulations of systems in contact with a particle reservoir (mentioned in \cref{subsec:Heff}) that was justified in Ref. \cite{DelleSite2024jpa} provides an indirect proof of the validity of the modified Lindblad equation \eqref{eq:modifiedLindblad}: in path integral molecular dynamics, the molecular trajectories are used to sample the quantum density operator of the system, thus indirectly solving the Lindblad equation for a system of quantum molecules, see, for example, Ref. \cite{Tuckerman2023}.
\end{remark}

\subsection{Computational protocol associated with the hierarchy}

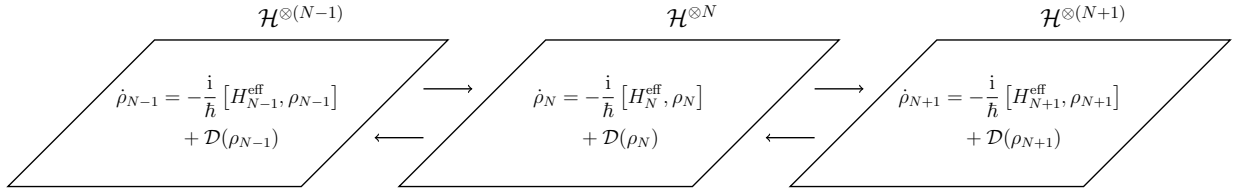
\begin{figure}
    \centering
    \scalebox{0.65}{
    \begin{tikzpicture}[font=\large]
        \draw[thick, xslant=1] (0,0) rectangle (6,3);
        \draw[thick, xslant=1] (8,0) rectangle (14,3);
        \draw[thick, xslant=1] (16,0) rectangle (22,3);
    
        \node at (6,3.5) {$\mathcal{H}^{\otimes (N-1)}$};
        \node at (14,3.5) {$\mathcal{H}^{\otimes N}$};
        \node at (22,3.5) {$\mathcal{H}^{\otimes (N+1)}$};
    
        \node[font=\normalsize] at (4.5,1.5) {%
            $\begin{aligned}
                \dot{\rho}_{N-1} &= - \frac{\mathrm{i}}{\hbar} \, \bigl[H_{N-1}^\mathrm{eff}, \rho_{N-1}\bigr] \\[2pt]
                                 &\quad + \mathcal{D}(\rho_{N-1})
            \end{aligned}$};
        \node[font=\normalsize] at (12.5,1.5) {%
            $\begin{aligned}
                \dot{\rho}_N &= - \frac{\mathrm{i}}{\hbar} \, \bigl[H_N^\mathrm{eff}, \rho_N\bigr] \\[2pt]
                             &\quad + \mathcal{D}(\rho_N)
            \end{aligned}$};
        \node[font=\normalsize] at (20.5,1.5) {%
        $\begin{aligned}
            \dot{\rho}_{N+1} &= - \frac{\mathrm{i}}{\hbar} \, \bigl[H_{N+1}^\mathrm{eff}, \rho_{N+1}\bigr] \\[2pt]
                             &\quad + \mathcal{D}(\rho_{N+1})
        \end{aligned}$};
    
        \draw[->, thick] (8.5,2) -- (9.5,2);
        \draw[->, thick] (8.5,1) -- (7.5,1);
        \draw[->, thick] (16.5,2) -- (17.5,2);
        \draw[->, thick] (16.5,1) -- (15.5,1);
    \end{tikzpicture}}
    \caption{Illustration of the sampling over $N$-space for the $N$-hierarchy \eqref{eq:modifiedLindblad} of Lindblad equations.}
    \label{fig:sampling}
\end{figure}

Equation \eqref{eq:modifiedLindblad} provides an $N$-hierarchy of equations for the full density operator of the system. The algorithm associated with this hierarchy for computing expectation values of observables can be described as follows (see Fig. \ref{fig:sampling} for a schematic illustration). Suppose that a specific realization of the system $\SFS$ with $N_0$ particles is given at time $t = t_0$. All the equations of the hierarchy \eqref{eq:modifiedLindblad} for $\rho_N$, in principle from $N = 1$ to the limit $N \to \infty$, can run in parallel to compute the solutions $\rho_N(t)$ for all $N \in \N$ and $t \in \R$. (Of course, this is an idealized picture; see below for a more practical version.) One now wants to determine the time series of the $\rho_N$, that is, the evolution of the system in terms of the number of particles $N$ (\enquote{effective time evolution}). At time $t = t_0$, where the system contains $N_0$ particles and it is assumed that in the next time step, the system will perform a single particle jump either to $N_0 + 1$ or $N_0 - 1$ total particles, one therefore considers the solutions $\rho_{N_0+1}$ and $\rho_{N_0-1}$. With these given density operators, one computes the ratios
\begin{equation}
    \frac{\tr(\rho_{N_0})}{\tr(\rho_{N_0+1})} \quad \text{and} \quad \frac{\tr(\rho_{N_0})}{\tr(\rho_{N_0-1})} \ ,
    \label{metropolis}
\end{equation}
where the traces are not computed over the full Fock space, but only over the $N$-particle subspaces, $N \in \set{N_0, N_0 + 1, N_0 - 1}$. Taking the minimum of these two ratios and comparing it to a threshold of acceptance (Metropolis criterion), one takes $\rho_{N_0 \pm 1}$, the choice of sign depending on the result of the previous step, as the new density operator if the move is accepted, or one keeps $\rho_{N_0}$ if the moves is rejected. Next, one proceeds to the time $t = t_0 + \Delta t$, $\Delta t > 0$. This procedure is repeated $K$ times, thus creating a chain
\begin{equation*}
    \rho_{N_0}(t), \ \rho_{N_0}(t_0+\Delta t) \ \text{or} \ \rho_{N_0 \pm 1}(t_0 + \Delta t), \ \dotsc, \ \rho_{N_0 \pm M}(t + K \Delta t)
\end{equation*}
with $M \in \set{0, 1, \dotsc, K}$. If $A$ is any observable of the system $\SFS$, and if $\mathfrak{N} = \{N_0, N_0 \pm 1, \dotsc, N_0 \pm M\}$ denotes the set of particle numbers obtained previously, then the expectation value of $A$ over the trajectory of system configurations computed before is given by
\begin{equation*}
    \braket{A} = \frac{1}{K} \sum_{N \in \mathfrak{N}} \tr(\rho_N A) \ .
\end{equation*}
In essence, the jump across the family of system configurations from one $N$-value to another occurs according to the instantaneous Boltzmann weight of close configurations.

This protocol relies on the idealized assumption that the solutions $t \mapsto \rho_N(t)$ of \cref{eq:modifiedLindblad} have been computed for every $N \in \N$; in practice, this is of course not realizable. However, the ideal algorithm described before can easily be adapted to be implementable in a concrete setup. Namely, one can start with an educated guess for the initial particle number $N_0$, e.g., the average $\braket{N}$ of equilibrium. Knowing that the distribution of $N$ for a weak coupling scenario is close to a normal distribution, one can focus the interest on a certain region around $\braket{N}$, say $[\braket{N} - \Delta N, \braket{N} + \Delta N]$, $\Delta N > 0$, and compute in parallel from the hierarchy \eqref{eq:modifiedLindblad} only the density operators $\rho_M$ for $M \in \set{\braket{N} - \Delta N, \dotsc, \braket{N} + \Delta N}$. As before, employing a Metropolis-like algorithm generates a time series of density operators with which one can compute averages of observables.

Regarding the complementarity with other models discussed in the Introduction, it is important to note that our derivation provides mathematical and physical consistency to the grand canonical sampling algorithms of Refs. \cite{Guimaraes2016, BulnesCuetara2016}. In fact, in equilibrium it automatically uses the eigenstates of the Hamiltonian $H_N - \mu N \id_S$, as manually done in Ref. \cite{Guimaraes2016}, and it also automatically employs the Metropolis-like algorithm of Eq. \eqref{metropolis} that was envisaged in Ref. \cite{BulnesCuetara2016} for climbing/descending on the $N$-ladder.

\section{Conclusions}\label{sec:conclusions}

We have analyzed the Lindblad equation in the context of the grand canonical ensemble, that is, in the limit of a stationary density operator in equilibrium with a large particle reservoir. We have put forward an analysis that underlines the difficulties which the Lindblad equation faces when used for modeling systems in equilibrium with a reservoir of particles. Current studies working with the Lindblad equation in a grand canonical ensemble require additional hypotheses to manually fit the equation into this scenario. Although numerically efficient, such approaches imply a degree of empiricism in a procedure that can be derived from first principles and thus be made self-contained.

Therefore, this paper suggests accepting the fact that the present derivation of the Lindblad equation for systems exchanging matter with an environment in (or close to) equilibrium cannot straightforwardly be justified. To overcome this problem, we have proposed a revision of the standard derivation, based on recent rigorous studies of grand canonical systems, that essentially relies on a new model \eqref{eq:revisedHint} for the system--reservoir interaction Hamiltonian which is founded upon physical approximations justified from general principles of statistical mechanics. The resulting modified Lindblad equation \eqref{eq:modifiedLindblad} automatically yields the correct density operator in the limiting case of stationary equilibrium without forcing the Lindblad coupling operator \eqref{eq:LindbladDissPart} into the conditions \ref{enu:cond1}, \ref{enu:cond2}, or \ref{enu:cond3} introduced in \cref{subsec:analysisCouplingOp}. Our approach automatically includes, and also mathematically justifies, the empirical impositions of other approaches, thus offering an effective generalization of the treatment of the Lindblad equation in a grand canonical ensemble.

We do not, however, exclude other possible approaches based on the idea of considering the density operator $\rho$ in the full Fock space, where the instantaneous particle numbers $N$ are the occupied states, varying as the system evolves. This particular path may be explored in future work; in this paper, we intended to propose a solution based solely on statistical mechanics considerations, which may help to build an efficient numerical algorithm based on the hierarchy \eqref{eq:modifiedLindblad}.



\acknowledgments{This work was supported by the DFG Collaborative Research Center 1114 ``Scaling Cascades in Complex Systems,'' Project No. 235221301, Project C01 ``Adaptive coupling of scales in molecular dynamics and beyond to fluid dynamics,'' and by the DFG, Project No. DE 1140/15-1, ``Mathematical model and numerical implementation of open quantum systems in molecular simulation.''}

\bibliography{lind.bib}

\end{document}